\documentclass[prd,showpacs,amsmath,
twocolumn,floatfix,amssymb, preprintnumbers, nofootinbib, superscriptaddress]{revtex4} 
\usepackage{hyperref}
\usepackage{epsfig,dcolumn}
\usepackage{graphicx}
\usepackage{comment} 
\usepackage{verbatim}
\usepackage{color}
\usepackage{graphicx}
\usepackage{bm}
\usepackage{ifpdf}
\usepackage{multirow}
\usepackage{soul} 

\def\lsim{\mathrel{\rlap{\lower4pt\hbox{\hskip1pt$\sim$}}
    \raise1pt\hbox{$<$}}}
\def\gsim{\mathrel{\rlap{\lower4pt\hbox{\hskip1pt$\sim$}}
    \raise1pt\hbox{$>$}}}
    
\begin{document}

\title{Three-body final state interaction in $\eta \to 3 \pi$ updated}
\author{P.~Guo}
\affiliation{Department of Physics and Engineering, California State University, Bakersfield, CA 93311, USA}
\author{I.~V.~Danilkin}
\email{danilkin@uni-mainz.de}
\affiliation{Institut f\"ur Kernphysik and PRISMA Cluster of Excellence, 
Johannes Gutenberg Universit\"at, D-55099 Mainz, Germany}
\affiliation{SSC RF ITEP, Bolshaya Cheremushkinskaya 25, 117218 Moscow, Russia}
\author{C.~Fern\'andez-Ram\'{\i}rez}
\affiliation{Instituto de Ciencias Nucleares, Universidad Nacional Aut\'onoma de M\'exico, Ciudad de M\'exico 04510, Mexico}
\author{V.~Mathieu}
\affiliation{Center for Exploration of Energy and Matter, Indiana University, Bloomington, IN 47403, USA}
\affiliation{Physics Department, Indiana University, Bloomington, IN 47405, USA}
\author{A.~P.~Szczepaniak}
\affiliation{Center for Exploration of Energy and Matter, Indiana University, Bloomington, IN 47403, USA}
\affiliation{Physics Department, Indiana University, Bloomington, IN 47405, USA}
\affiliation{Theory Center, Thomas Jefferson National Accelerator Facility, Newport News, VA 23606, USA}
 
\collaboration{Joint Physics Analysis Center}
\preprint{JLAB-THY-16-2302}

\date{\today}

\begin{abstract} 

In view of the recent high-statistic KLOE data for the $\eta \to \pi^+ \pi^- \pi^0$ decay, a new determination of the quark mass double ratio has been done. Our approach relies on a dispersive model that takes into account rescattering effects between three pions via subenergy unitarity. The latter is essential to reproduce the Dalitz plot distribution. A simultaneous description of the KLOE and WASA-at-COSY data is achieved in terms of just two real parameters. From a global fit, we determine $Q=21.6 \pm 1.1$. 
The predicted slope parameter for the neutral channel $\alpha=-0.025\pm 0.004$ is in reasonable agreement with the PDG average value.

\end{abstract} 



\maketitle

\section{Introduction}\label{intro}

Three meson systems play an important role in studies of hadron reaction dynamics and hadron spectroscopy. For example, in three-particle  decays of heavy quarkonia  several candidates for non-quark model resonances have recently been observed  \cite{PDG-2015, Aaij:2014jqa, Swanson:2006st}. Three-body decays of $B$ and $D$ mesons are a promising laboratory for studies of  CP-violation  \cite{Aaij:2013sfa,Aaij:2013bla}. In the light meson sector the limited phase space makes three-particle decays an ideal testing ground of effective theories of strong interactions. Detailed amplitude analysis of three meson production  becomes even more  important in light of the current and forthcoming high precision data  from various hadron facilities \cite{Battaglieri:2010zza,Eugenio:2003,Adolph:2014rpp, Ablikim:2015cmz}.

The isospin breaking $\eta\rightarrow 3\pi$ decay, which we consider here, is of great importance as it allows to measure the light quark mass difference. The electromagnetic effects are known to be small \cite{Sutherland:1966zz,Bell:1996mi,Ditsche:2008cq}, and the decay is  driven by strong interactions through the $\Delta I=1$ isospin breaking transition that appears directly in the QCD Lagrangian. The decay amplitude is proportional to the light quark mass difference, $(m_u-m_d)$, and it is conventionally expressed in terms of the parameter $Q^2$ defined by 
\begin{equation}\label{Eq:Q}
\frac{1}{Q^2}=\frac{m_d^2-m_u^2}{m_s^2-\hat m^2}\,,\quad \hat m =\frac{(m_u+m_d)}{2}\,,
\end{equation}
with $m_s$ being the strange quark mass. Note, that this double ratio (\ref{Eq:Q}) is protected to strong high order corrections. Given the small breakup momenta  the distribution of pions in the Dalitz plot  of the $\eta\rightarrow \pi^+\pi^-\pi^0$ decay can be analyzed in terms of a small number of parameters that determine deviations from a uniform distribution. These parameters are referred to as Dalitz plot parameters. Early analyses \cite{Gormley:1970qz,Layter:1973ti,Abele:1998yj} could determine only a few, leading Dalitz plot paramters. A few more parameters were determined using the 2008 KLOE measurement \cite{Ambrosino:2008ht}. These analyses were further improved thanks to the high-quality WASA-at-COSY  \cite{Adlarson:2014aks}  and new KLOE \cite{Anastasi:2016qvh} data.  The statistics  of the most resent measurements is high enough to allow for binned, data-driven analysis.

On the theoretical side there has been significant progress in chiral, effective field theory analysis of $\eta$ decays. Chiral perturbation theory ($\chi$PT) seems to converge poorly, yielding $\Gamma_{\eta\rightarrow \pi^+\pi^-\pi^0}=$ 66, $167\pm50$, $\sim 300$ eV at leading (LO), next-to-leading (NLO) and next-to-next-to-leading (NNLO) order, respectively \cite{Cronin:1967jq, Osborn:1970nn, Gasser:1984pr, Bijnens:2007pr}. This indicates importance of pion-pion interactions and various approaches have been used to implement these effects to all orders \cite{Colangelo:2009db,Lanz:2013ku, Schneider:2010hs, Kampf:2011wr, Descotes-Genon:2014tla}.

In our recent study \cite{Guo:2015zqa}  we implemented  the S-matrix constraints of unitarity, analyticity and crossing symmetry via a set of dispersion relations \cite{Khuri:1960kt, Kambor:1995yc, Anisovich:1996tx}. Connection with QCD is achieved by  matching the dispersive amplitudes with $\chi$PT at the point where the latter converges best {\it  i.e.}  below the threshold. In  \cite{Guo:2015zqa} we used WASA-at-COSY data \cite{Adlarson:2014aks} to determine the free parameters, {\it i.e.}  subtraction constants of the dispersive integrals. As a result, we achieved a simultaneous description of the Dalitz plot distributions of the charged and neutral $\eta$ decay modes. In order to extract the parameter  $Q$ we matched the dispersive amplitude with the next-to-leading order (NLO) $\chi$PT result near the Adler zero and we obtained $Q=21.4\pm0.4$ \cite{Guo:2015zqa}. The purpose of this letter is to revisit the result of \cite{Guo:2015zqa} in a view of the new high statistic data from the KLOE experiment \cite{Anastasi:2016qvh}.

\section{The method}

In this section, we briefly review the $\eta$ decay amplitudes that were developed in \cite{Guo:2015zqa}. For $\eta\to 3\pi$, the transition amplitude $A(s,t,u)$ is a function of three Mandelstam variables \mbox{$s=(p_{\pi^+} + p_{\pi^-})^2$}, \mbox{$t=(p_{\pi^-} + p_{\pi^0})^2$}, and \mbox{$u = (p_{\pi^+} + p_{\pi^0})^2$} which are related by \mbox{$s+t+u = m_\eta^2 + 3\,m_{\pi}^2$}. Except for the phase-space boundary, we work in the isospin limit and take $m_{\pi}  =  (2\,m_{\pi^{+}}+m_{\pi^{0}})/3$. At low energies, one can perform a partial wave (p.w.) decomposition while crossing symmetry implies unitarity cuts in all three Mandelstam variables. Therefore we symmetrize the p.w. expansion in all three channels \cite{Khuri:1960kt,Bronzan:1963kt,Aitchison:1965kt,Aitchison:1965dt,Aitchison:1966kt,Pasquier:1968kt,Pasquier:1969dt}, which for the charged decay, $\eta\to \pi^+\pi^-\pi^0$, implies the following representation, 
\begin{align} 
& A^{C}(s,t,u) =     \sum_{L=0}^{L_{max}=1}\frac{(2L+1)}{2} \bigg [ \frac{2}{3}\,P_{L}(z_{s})  \left ( \, a_{0 L}  (s ) - a_{2 L} (s) \right)  \nonumber \\
& + P_{L}(z_{t})  \left ( \,a_{1 L} (t) +  a_{2 L} (t) \right)-P_{L}(z_{u})  \left ( \,a_{1 L} (u) - a_{2 L} (u) \right)  \bigg ]\,.\nonumber \\
\label{ampc}
\end{align}
The amplitudes $a_{IL}$ have only the right-hand, unitary cuts. In Eq. (\ref{ampc}) $z_{i}\equiv\cos\theta_i$ and $\theta_{s,t,u}$ are the center-of-mass scattering angles in the $s$, $t$ and $u$-channels, respectively. The subscript \mbox{$(I,L)$} labels isospin and orbital angular momentum, with \mbox{$I+L=even$} due to Bose symmetry. The latter implies that for $L_{max}=1$ there are three unknown isospin amplitudes with $(I,L)=(0,0),\,(2,0),\,(1,1)$. An amplitude of the neutral decay, $\eta\to 3\pi^0$ can be easily reconstructed using $\Delta I=1$ relation, 
\begin{equation}\label{ampn}
A^{N}(s,t,u) =A^{C}(s,t,u) +A^{C}(t,u,s) +A^{C}(u,s,t)\,,
\end{equation}
We emphasize, that the decomposition in Eq.(\ref{ampc}) has the same analytical properties as the amplitude in  NNLO chiral expansion \cite{Stern:1993rg, Knecht:1995tr}. However, in contrast to $\chi$PT, we can impose unitarity to all orders on the  $a_{IL}$ amplitudes. The discontinuity along the right-hand cut can be expressed through the elastic $\pi\pi$ partial wave amplitudes $f_{I  L}$, which leads to
\begin{align}\label{dis}
\Delta a_{I  L} (s)  &=\frac{1}{2\,i}\left(a_{IL}(s+i\epsilon)-a_{IL}(s-i\epsilon)\right)\nonumber\\
&= f_{I  L}^{*}(s) \rho(s) \bigg( a_{I L} (s)+2\sum_{L'=0}^{L_{max}}\sum_{I'}(2L'+1)  \nonumber \\
&\quad \times\int_{-1}^{+1}\frac{dz_s}{2}\,P_{L}(z_{s})\,P_{L'}(z_{t}) C_{st}^{I  I'}  a_{I' L'} (t)   \bigg)\,,
\end{align}
Note, that the amplitudes $a_{I L} (s)$ and $f_{I  L}(s) $ for $L>0$ are subject to kinematical constraints  \cite{Lutz:2011xc} which have to be removed before application of  dispersion relations. This is done by introducing the reduced amplitudes $\tilde a_{I L} (s)= a_{IL}(s)/Z_L(s)$ where the factor $Z_L(s)$ is proportional to the product of the c.m. momenta of $\pi\pi$ and $\pi \eta$ \cite{Guo:2015zqa}. The normalization is fixed by the phase space factor $\rho(s) =\sqrt{1- 4\,m_{\pi}^{2}/s}$, with  \mbox{$\mbox{Im} (1/f_{I L}(s)) =-\rho(s)$}. The explicit form of the crossing matrices $C_{st, su}^{I  I'}$ can be found in \cite{Guo:2015zqa}. The contribution from the first term on the right-hand side of Eq.\,(\ref{dis}) reproduces the direct $s$-channel unitarity, while the second term contains the left-hand cuts from the $t$ and $u$-channels. While calculating the latter, special care has to be taken for \mbox{$4\,m_\pi^2 \leq s <(m_\eta+m_\pi)^2$}, i.e. one has to deform the contour to avoid the cut along the real axis \cite{Gribov:1962fu,PhysRev.132.2712,Bronzan:1963kt}. The kinematical singularity free amplitudes  $\tilde{a}_{IL}(s)$ satisfy the Cauchy representation, which up to subtraction constants yields, 
\begin{equation} 
\tilde{a}_{IL}(s) = \frac{1}{\pi} \int_{4m_{\pi}^2}^\infty \, ds' \frac{\Delta \tilde{a}_{IL}(s')}{s' -s}\,.\label{dis2} 
\end{equation}
The combination of (\ref{dis2}) and (\ref{dis}) sets the so-called Khuri-Treiman (KT) framework that can be solved using  several techniques \cite{Guo:2014vya}. The most popular method is to write a set of dispersion relation for the ratio $\tilde{a}_{IL}(s)/\Omega_{IL}(s)$ with $\Omega_{IL}(s)$ being the Omn\`es function \cite{Colangelo:2009db,Descotes-Genon:2014tla}. In this case it is necessary to make further assumptions about the unknown high-energy region which is typically done by introducing subtractions. This procedure is relatively easy for $\omega\rightarrow 3\pi$ decay which depends dominantly on the pion-pion P-wave scattering input \cite{Niecknig:2012sj,Danilkin:2014cra}. However, this is more challenging for $\eta\to 3\pi$ decay, where the dominant contribution comes from the S-wave and the form of the $\pi\pi$ isoscalar Omn\`es function is very sensitive to  the asymptotic behavior of the phase shift $\delta_{IL=00}(s\to \infty)$. In \cite{Colangelo:2015kha} (Figs. 4 and 7) different scenarios for the the $\pi\pi$ phase shift inputs were investigated and the corresponding Omn\`es functions were produced. In order to minimize these differences, the subtraction polynomial of the sufficient  order is required in the dispersion representation.

A complementary approach is the Pasquier inversion \cite{Pasquier:1968kt,Aitchison:1978pw} that we applied to analyze WASA-at-COSY data in \cite{Guo:2015zqa} and use here as well. This method uses contour deformation to exchange the order of double integral appearing on the right hand side of Eq.~(\ref{dis2}). As it was shown in \cite{Guo:2014vya}, once the two-body amplitudes $f_{IL}(s)$ are known, different methods for solving the disersive integral provide the same result. However, when the Pasquier inversion is applied, the input of $f_{IL}(s)$  is required in a different energy region.

To proceed we write $\tilde{a}_{IL}(s)$ in the form 
\begin{equation}\label{Eq:product}
\tilde{a}_{IL}(s) =  \mathcal{F}_{IL}(s) \,f_{IL}(s)  \,g_{IL}(s),
\end{equation}
where the function $\mathcal{F}_{IL}(s)$ is introduced to remove Adler zeros specific to the elastic amplitude $f_{IL}(s)$ and introduce zeros in the decay amplitude as required by chiral symmetry. From Eq.~(\ref{dis}) and Eq.~(\ref{Eq:product}) one can derive the  discontinuity of $g_{IL}$ and write  the dispersive representation for $g_{IL}(s)$. As a result we obtain a double integral equations for \mbox{$g_{IL}(s)$}, which can be reduced to a single integral equation using the Pasquier inversion,
\begin{align}
g_{IL}(s)  = &  -\frac{1}{\pi} \int_{-\infty}^{ 0} ds' \frac{1}{s' - s}  \frac{\Delta f_{IL}(s')}{f_{IL}^{*}(s')}\,g_{IL}(s')   \nonumber \\
&+ \frac{1}{\pi} \int_{-\infty}^{(M-m_{\pi})^2}  dt\sum_{L'=0}^{L_{max}}\sum_{I'} \mathcal{K}_{ I L, I' L'}(s,t)  \nonumber \\
& \quad \quad \quad \quad  \quad \quad  \quad    \times   C_{st}^{I  I'}   f_{I' L'}(t)\,g_{I'L'}(t). \label{pasqg}
\end{align} 
The explicit form of the kernel functions,  \mbox{$\mathcal{K}_{IL, I' L'}(s,t)$} can be found in \cite{Guo:2015zqa}. Currently, they are only calculated in the region $s\in (0,\,(M-m_{\pi})^2)$, which includes physical region and therefore cover the whole Dalitz plot region. In order to compute the amplitudes beyond that region, a proper analytical continuation is required within Pasquier inversion technique. Important steps in that direction were already elaborated in \cite{Guo:2014mpp}. We will come this issue later in the $Q$-value determination.

The first term and the part of the second term on the right-hand side have the left-hand cut and can be expanded in the Taylor series in the physical region. Retaining only a single term in the expansion we arrive at the following relation \cite{Guo:2014vya, Guo:2015kla}
\begin{align}\label{gapprox}
&g_{IL}(s)  =   g_{IL}(s_0) + \frac{1}{\pi} \int_{0}^{(M-m_{\pi})^2} dt \sum_{L'=0}^{L_{max}}\sum_{I'}  
C_{st}^{I  I'}  \nonumber \\
 &\quad \quad \times \big(\mathcal{K}_{IL,I'L'}(s,t) -\mathcal{K}_{IL,I'L'}(s_{0},t)\big)  f_{I' L'}(t)\,g_{I'L'}(t)\,,
\end{align}
which is solved by discretizing the integral and inverting the kernel matrix. The subtraction point, $s_0$  is chosen to be near the Adler zero at leading order of $\chi$PT, \mbox{$s_0 \simeq 4/3\,m_\pi^2$}, and the subtraction constants $g_{IL}(s_0)$, which absorb the left hand cut contribution  are the free parameters that are to be determined by fitting to the data.

\section{Numerical results} 
\label{SectionIII}

\begin{table}[t]
\centering
\caption{Results of two-body (2b) and three-body (3b) fits to  WASA-at-COSY \cite{Adlarson:2014aks} data, KLOE data \cite{Anastasi:2016qvh} and the combined fit. For two-body fits we quote $(g^{2b}_{IL}(s_0) \pm \Delta g^{2b}_{IL}(s_0))/g^{2b}_{00}(s_0)$, while when presenting results of three-body fit we quote $(g^{3b}_{IL}(s_0) \pm \Delta g^{3b}_{IL}(s_0))/g^{2b}_{00}(s_0)$, where $g^{2b}_{00}(s_0)$ is the central value obtained in the two-body fit with the same number of partial waves. We do the latter to illustrate the relative change in normalization between two- and three-body fits. 
\label{tab:par}}
\renewcommand{\arraystretch}{1.8}
\fontsize{8}{2}
\begin{tabular*}{0.5\textwidth}{@{\extracolsep{\fill}}lcccr@{}}%
\hline\hline
\multicolumn{5}{l}{$(I,L)=(0,0),\,(1,1)$} \\
\hline\hline
&    $g_{00}/g_{00}^{(2b)}$ &  $g_{20}/g_{00}^{(2b)}$ & $g_{11}/g_{00}^{(2b)}$ & \mbox{$\chi^{2}/d.o.f.$}\\
\hline\hline
(2b)   &&&  &  \\
 COSY  & $1.000 \pm 0.002$        & --          &  $0.058 \pm 0.009$& $1.45$ \\
KLOE  & $1.000 \pm 0.005$        & --          &  $0.019 \pm 0.025$& $10.4$ \\
Comb &   $1.000 \pm 0.005$     & --          &  $0.020 \pm 0.026$   & $9.5$ \\
\hline
(3b) && & &\\
 COSY  &$1.043 \pm 0.005$& -- & $0.233 \pm 0.009$& $0.95$ \\
KLOE  &$1.046 \pm 0.006$& -- & $0.194 \pm 0.024$& $2.61$ \\
Comb  &   $1.046 \pm 0.006$    & --          & $0.195 \pm 0.026$  & (Set 1) $1.64$\\

\hline\hline
\multicolumn{5}{l}{$(I,L)=(0,0),\,(2,0),\,(1,1)$} \\
\hline\hline
&    $g_{00}/g_{00}^{(2b)}$ &  $g_{20}/g_{00}^{(2b)}$ & $g_{11}/g_{00}^{(2b)}$ & \mbox{$\chi^{2}/d.o.f.$}\\
\hline
(2b) && & &\\
 COSY  & $1.00 \pm 0.02$& $-0.26\pm 0.05$ & $0.38\pm 0.07$ & $0.94$
\\
KLOE & $1.00 \pm 0.06$& $-0.44\pm 0.17$ & $0.56\pm 0.21$ & $1.21$
\\
Comb &  $1.00 \pm 0.07$     &$-0.44 \pm 0.18$         &  $0.56 \pm 0.23$  & $1.54$  \\
\hline
(3b) && & &\\
COSY  &$1.19 \pm 0.01$        & $0.14\pm 0.003$ & $0.28 \pm 0.04$ & $0.90$\\
KLOE &$1.215 \pm 0.002$        & $0.015\pm 0.005$ & $0.427 \pm 0.008$ & $1.29$\\
Comb &  $1.214 \pm 0.002$     & $0.018 \pm 0.005$          &  $0.423 \pm 0.008$ &(Set 2) $1.61$\\

\hline\hline
\end{tabular*}
\end{table}

\begin{table*}[t]
\centering
\caption{Dalitz plot parameters for \mbox{$\eta \to \pi^{+} \pi^{-} \pi^{0}$}. We present the results of the seprate fits to WASA-at-COSY \cite{Adlarson:2014aks} and KLOE \cite{Anastasi:2016qvh} data and the combined fit. In both cases we average Dalitz Plot paramteres between $(I,L)=(0,0),\,(1,1)$ and $(I,L)=(0,0),\,(2,0),\,(1,1)$ scenarios (see Table \ref{tab:par}). \label{tab:xy}}
\renewcommand{\arraystretch}{1.8}
\fontsize{8}{2}
\begin{tabular*}{\textwidth}{@{\extracolsep{\fill}}llllll@{}}
\hline\hline
 &  $a$&    $b$ &  $d$ & $f$ & $g$ \\
\hline
%
WASA-at-COSY \cite{Adlarson:2014aks}      & $-1.144\pm0.018$ &      $0.219\pm0.019\pm0.037$ 
& $0.086\pm0.018\pm0.018$ & $0.115\pm0.037$  & -- \\
KLOE \cite{Anastasi:2016qvh}   & $-1.095\pm 0.003^{+0.003}_{-0.002}$ & $0.145\pm0.003\pm0.005$ & $0.081\pm0.003^{+0.006}_{-0.005}$ & $0.141\pm0.007^{+0.007}_{-0.008}$ & $-0.044\pm0.009^{+0.012}_{-0.013}$ \\
%
%
\hline\hline
Theory (fit to COSY)
&$-1.116\pm 0.032$ & $0.188\pm 0.012$ &
$0.063\pm 0.004$ & $0.091\pm 0.003$ & $-0.042 \pm 0.009$  \\
Theory (fit to KLOE) &$-1.077\pm 0.029$ & $0.170\pm 0.008$ &
$0.060\pm 0.002$ & $0.091\pm 0.003$ & $-0.044 \pm 0.003$  \\
Theory (combined fit) &$-1.075\pm 0.028$  &  $0.155\pm 0.006$ & $0.084\pm 0.002$ & $0.101 \pm 0.003$  &  $-0.074\pm 0.003$ \\ 
\hline\hline
\end{tabular*}
\end{table*}

The $\eta\to 3\pi$ Dalitz plot distribution is conventionally expressed in terms of the variables $x,y$ which are defined by 
\begin{align}\label{defXY}
x &= \frac{\sqrt{3}}{2\,m_\eta\,Q_{c}}\,(t-u)\,, \nonumber\\
y &= \frac{3}{2\,m_\eta\,Q_{c}} \left ((m_\eta-m_{\pi^{0}})^{2} -s \right ) -1\,.
\end{align}
For charge decay \mbox{$Q_{c} = m_\eta-2\,m_{\pi}^+-m_\pi^0$}  and for the neutral decay \mbox{$Q_{n} = m_\eta-3\,m_\pi^0$}. Kinematics restrict the events to be contained within the unit disk  $x^2+y^2\leq 1$. The KLOE \cite{Anastasi:2016qvh} and WASA-at-COSY \cite{Adlarson:2014aks} data were binned into 371 and 59 sectors of the unit disk,  respectively (only bins that lie completely inside the physical region are included). We determine the  unknown parameters, $g_{IL}(s_0)$ by minimizing
\begin{align}
\chi^{2} = \sum^{N}_{bins}  \left ( \frac{|A|_{\mbox{data}}^{2} - |A^{C}\left(\{g_{IL}(s_{0})\}\right)|^{2} }{\Delta |A|^2_{\mbox{data}} }\right)^{2}\,, \label{c2} 
\end{align}
where $|A|_{\mbox{data}}^2$ is the acceptance corrected data in each bin and $\Delta |A|_{\mbox{data}}$ is the uncertainty (assumed to be only statistical).

In our earlier work \cite{Guo:2015zqa} we performed two different fits to the WASA-at-COSY data. The first fit was done using only two resonant p.w. amplitudes, i.e. $(I,L)=(0,0),\,(1,1)$ and the second fit included all isospin amplitudes for the $S$ and the $P$ waves, i.e. $(I,L)=(0,0),\,(2,0),\,(1,1)$. Here we follow the same procedure. The resulting parameters are collected in Table \ref{tab:par}, where we show fits with and without three-particle rescattering effects (so called "two-body" and "three-body" scenarios), i.e using (\ref{gapprox}) to determine $g_{IL}(s)$, or just setting it to a constant $g_{IL}(s)=g_{IL}(s_0)$.

\begin{figure*}[t]
\includegraphics*[width=0.49\textwidth]{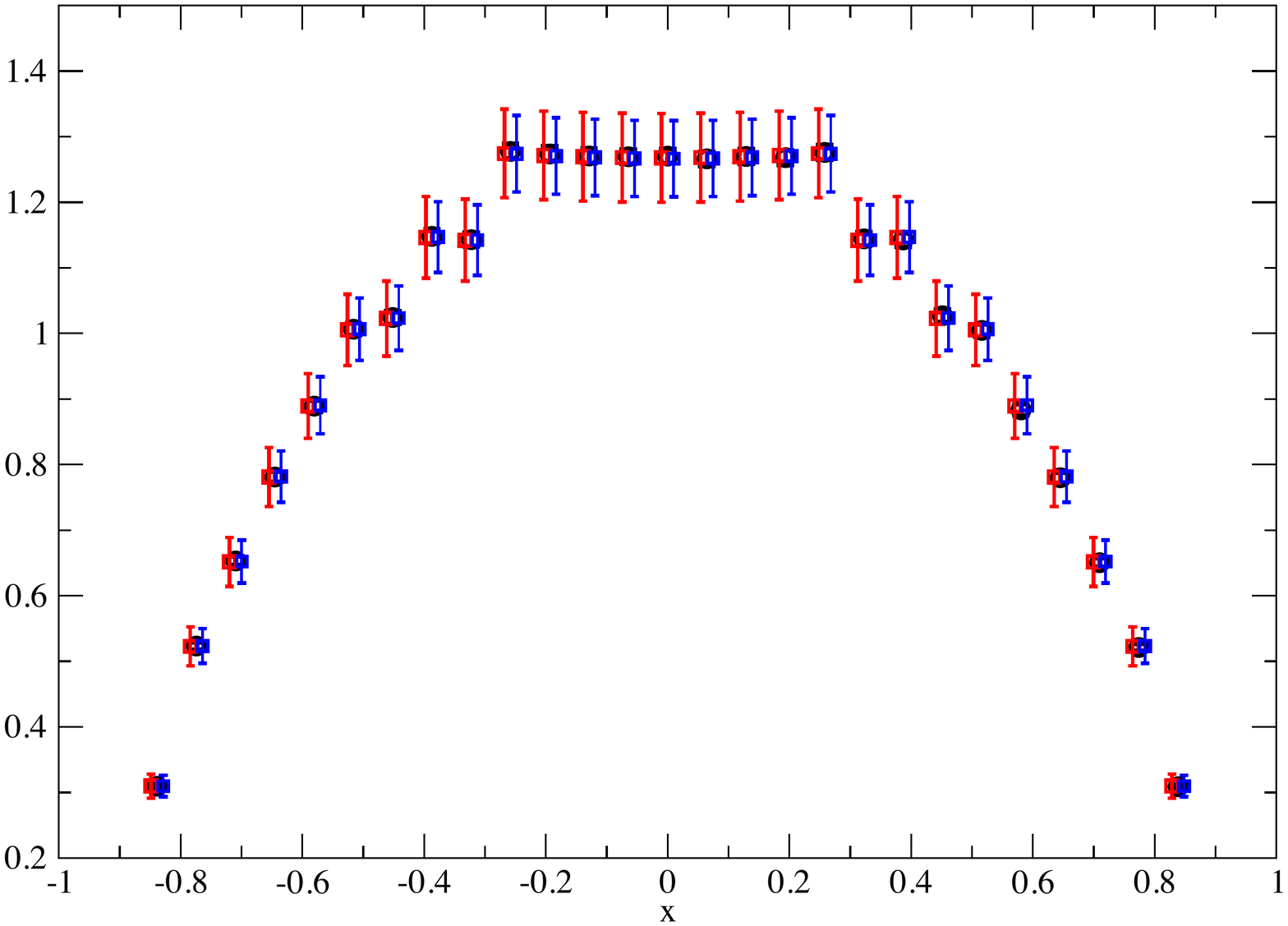}
\includegraphics*[width=0.49\textwidth]{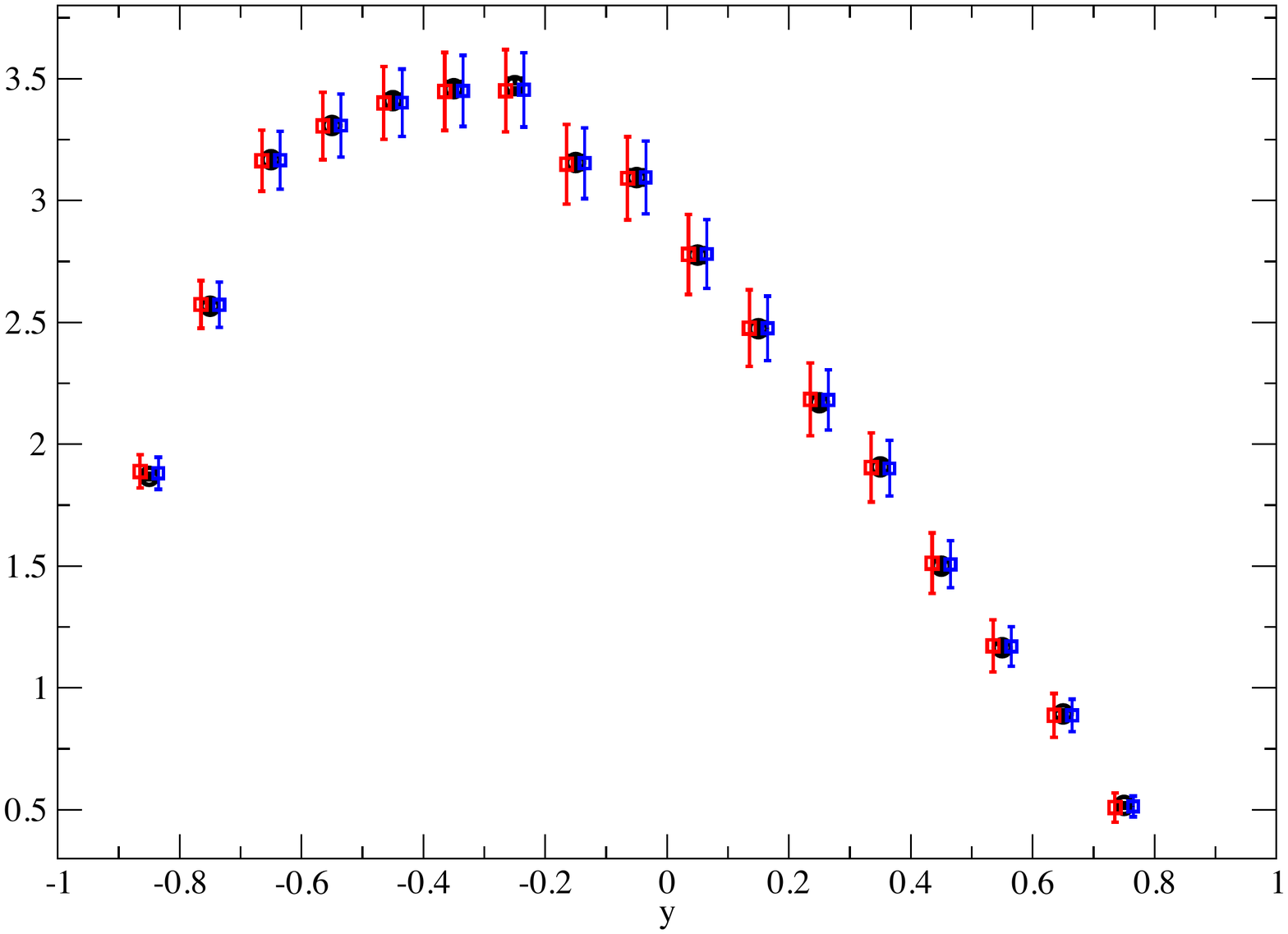}
\includegraphics*[keepaspectratio,width=0.49\textwidth]{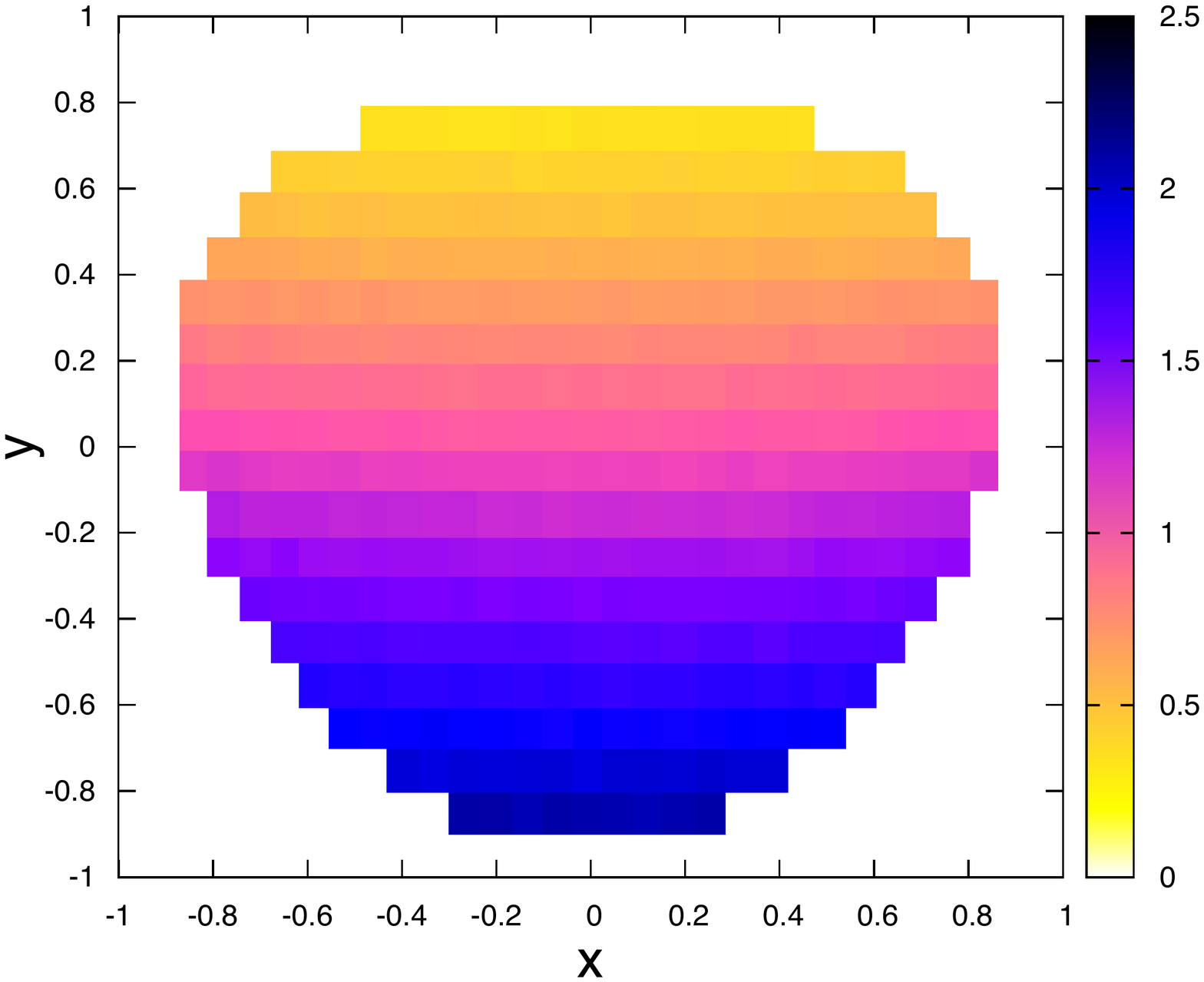}
\includegraphics[keepaspectratio,width=0.49\textwidth]{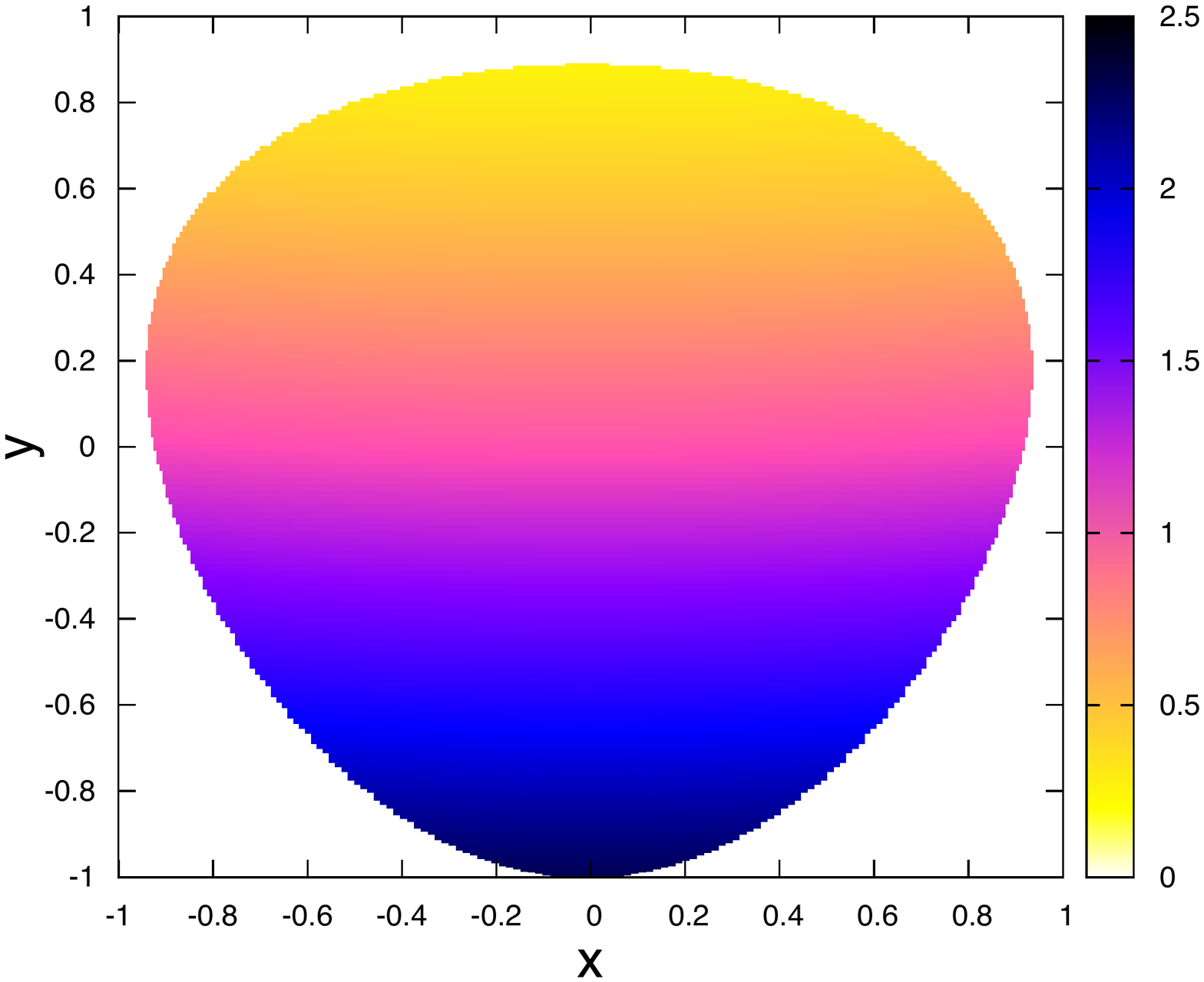}
\caption{Upper panels are the $x$- and $y$-projection plots.  Black circles are the data. Red squares and blue squares represent results of the  two-body and three-body fits, respectively. The fits are performed on the Dalitz distribution shown in the bottom left panel using three waves, \mbox{$(I,L)=(0,0),\,(2,0),\,(1,1)$}. For better visualization the experimental points are shifted horizontally from the fit results. The bottom right panel is the Dalitz distribution from the three-body fit with \mbox{$(I,L)=(0,0),\,(2,0),\,(1,1)$} waves, while the bottom left panel is the new KLOE \cite{Anastasi:2016qvh} data.
\label{fig:proj1}}
\end{figure*}

In the first step we fit the KLOE data alone. When only $(I,L)=(0,0),\,(1,1)$ amplitudes are taken into account,  we observe a significant reduction of $\chi^2/d.o.f$ while moving from the "two-body" to the "three body" case. At the same time, when a complete set of $S$ and $P$ waves is incorporated, the $\chi^2/d.o.f$ stabilizes at around 1.2-1.3 in both cases. In the second step, we combine the KLOE and WASA-at-COSY data. The results are in general very similar, showing the consistency of two different data sets. The results of the fit are shown in Fig. \ref{fig:proj1}.

The Dalitz plot parameters are defined as an effective range expansion around the center of the Dalitz plot $x=y=0$,
\begin{align}\label{Dalitzpar1}
\frac{|A^{C}(x,y)|^{2}}{|A^{C}(0,0)|^{2} } &=1+ a\,y + b\,y^{2}+ d\,x^{2 }+ f\,y^{3}+ g\,x^{2} y + \cdots \nonumber\\
\frac{|A^{N}(z,\phi)|^{2}}{|A^{N}(0,0)|^{2}}&= 1+ 2\,\alpha\,z + 2\,\beta\,z^{3/2}\,\sin 3 \phi  + \cdots\,,
\end{align}
where \mbox{$x=\sqrt{z} \cos \phi$} and \mbox{$y=\sqrt{z} \sin \phi$}. In Table \ref{tab:xy} we show the averaged Dalitz Plot parameters between  three-body fits with $(I,L)=(0,0),\,(1,1)$ and $(I,L)=(0,0),\,(2,0),\,(1,1)$ wave sets. We also predict the slope parameter $\alpha$ for the neutral decay mode to be
\begin{eqnarray}\label{alpha_new}
&&\alpha= -0.024  \pm0.004 ,\quad \beta=-0.000 \pm 0.002\,,\\
&&\alpha= -0.025  \pm0.004 ,\quad \beta=-0.000 \pm 0.002\,,\nonumber
\end{eqnarray}
from the KLOE and combined KLOE \& WASA-at-COSY fits, respectively. Note, that without three body effects $\alpha^{2b}= -0.021\pm0.004$ for both sets of data. The new results (\ref{alpha_new}) compares favorably with the most recent PDG value $\alpha^{PDG}=-0.0315\pm0.0015$ \cite{PDG-2015}. This difference is expected to get even smaller once electromagnetic corrections are fully considered (not only in kinematic factors).

\begin{figure*}[t]
\includegraphics*[width=0.9\textwidth]{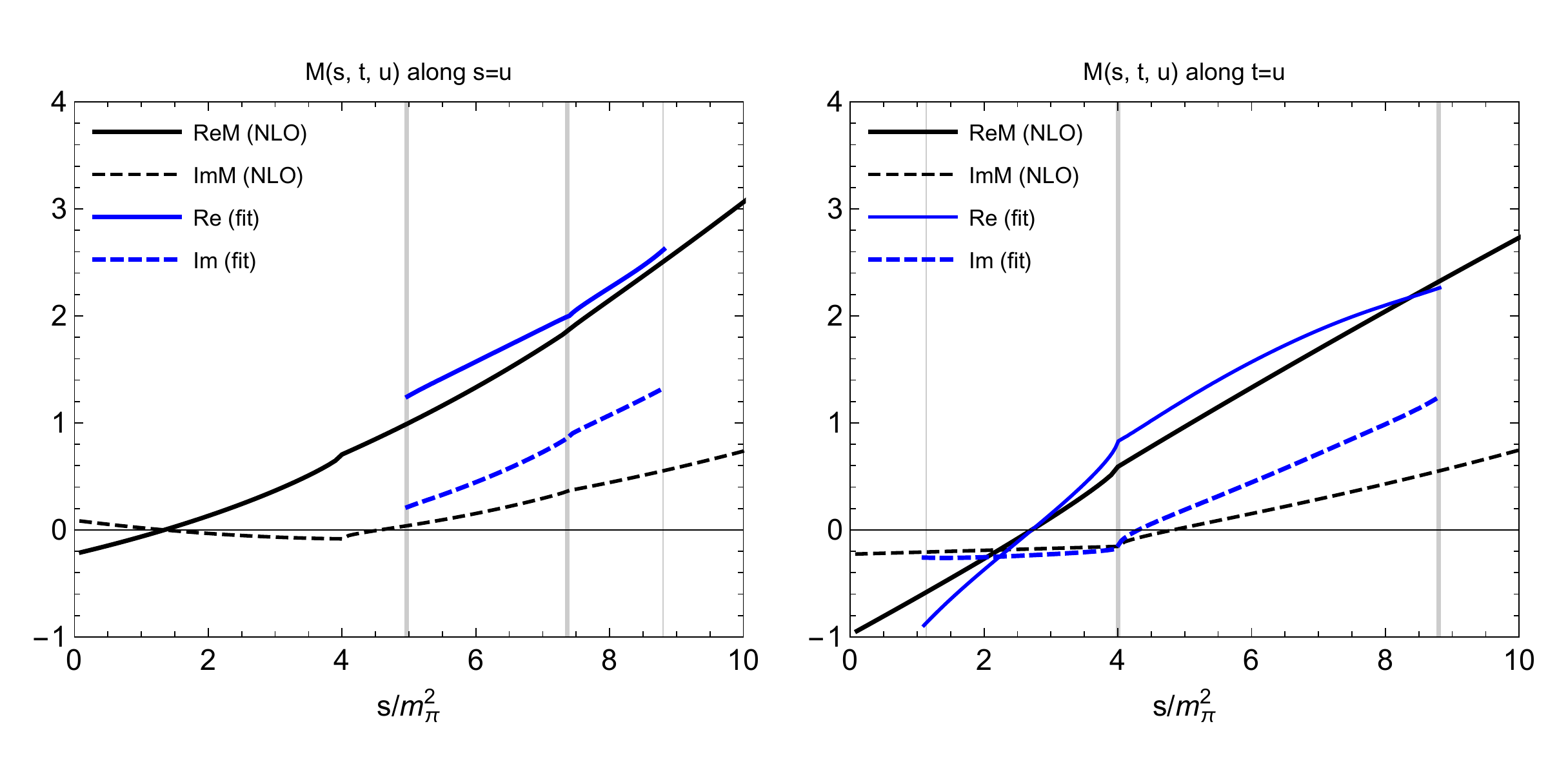}
\caption{The amplitude along the lines $s=u$ and $s=t$ with a comparison to NLO ChPT. The relation between $M(s,t,u)$ and $A^C(s,t,u)$ is given in \cite{Guo:2015zqa}. The solid thick and thin vertical lines correspond to the physical region and the region where we calculated our amplitudes, respectively. As explained in the text, in order to compute the amplitudes beyond that region, a proper analytical continuation is required within Pasquier inversion technique.}\label{fig:2}
\end{figure*}

\subsection{Matching to $\chi$PT and the $Q$-value}

We note, that NLO $\chi$PT result depends on four low energy constants (LECs). These can be reduced to a single one $L_3=(-2.35\pm0.37)\times 10^{-3}$ \cite{Amoros:2001cp} if one  employs Gell-Mann-Okubo constraint between meson masses and meson decay constants. This is  not the case at NNLO where one has to deal with several  unknown LEC's. Therefore, in our analysis we match the dispersive amplitudes with NLO $\chi$PT near the  Adler zero. Note, that we match single variable partial wave amplitudes $a_{IL}(s)$ to $\chi$PT and not the full amplitude $A^C(s,t,u)$ along the lines $s=t$ or $t=u$. This procedure should be equivalent, since $a^{\chi PT}_{IL}(s)$ possess Adler zeros as well. In order to perform the matching at $t=u$ we would need to make an additional analytic continuation of our results. In Fig. \ref{fig:2} we show our results of a combined fit to KLOE and WASA-at-COSY  with a fixed overall normalization to NLO $\chi$PT near the  Adler zero in the region $s=(0,\,10\,m_\pi^2)$ along the lines $s=t$ or $t=u$. The updated Q-value is
\begin{equation}\label{finalQ}
Q=21.6 \pm 1.1\,,
\end{equation}
which should be compared to the result of \cite{Guo:2015zqa} \mbox{$Q=21.4\pm 1.1$} (the fit to WASA-at-COSY data only) and \mbox{$Q=21.7 \pm 1.1$} (the fit to KLOE data only).
Note, that the obtained $Q$-value is consistent with the latest ($N_f= 2+1+1$) lattice computations $Q=22.2\pm1.8$ \cite{Aoki:2016frl}.

There are several challenges in the accurate determination of the $Q$-value. The first one comes from the elastic $\pi\pi$ scattering amplitudes, which are available from the Roy equation analysis \cite{GarciaMartin:2011cn} and implies the error $\Delta Q_{\pi\pi}=0.25$. Second uncertainty is due to experimental $\eta\rightarrow \pi^+\pi^-\pi^0$ decay width, which serves as an input in our analysis. Its value increased by more than 3$\sigma$ over the last thirty years, resulting in the current PDG value $\Gamma_{\eta\to\pi^+\pi^-\pi^0}=296\pm16$ eV \cite{PDG-2015}. This error propagates to $\Delta Q_{\Gamma}=0.29$. Third source of uncertainty is the experimental data on Dalitz plot itself, which thanks to the recent high-statistical analyses has improved significantly. Its contribution to the Q-value is $< 10\%$ of the size of error bars coming from the $\pi\pi$ amplitudes and therefore we included it in $\Delta Q_{\pi\pi}$. Another uncertainty comes from matching to $\chi$PT amplitude. The error associated with $L_3$ LEC is very small and therefore the resulting error bar in our previous analysis was $\Delta Q_{\text{total}}=0.4$ \cite{Guo:2015zqa}. That error was dominated by the experimental error bars and therefore should be viewed as a lower bound of the full error. We note, however, that the Q-value determination is very sensitive to the matching to NLO amplitude. Though the region around the Adler zero is supposed to be stable against contributions from higher orders in the chiral expansion, we cannot completely exclude them. Assuming conservatively an additional error of 10\% on NLO $\chi$PT amplitude, gives $\Delta Q_{match}=1.08$ and the total $\Delta Q_{\text{total}}=1.1$ quoted in Eq. (\ref{finalQ}).

\section{Conclusions}

In this work we revisited our previous dispersive analysis \cite{Guo:2015zqa} of the $\eta\rightarrow 3\pi$ decay in light of the new KLOE \cite{Anastasi:2016qvh} data. Within our unitary model we established a unified description of charged and neutral decay modes. The method is based on Khuri-Treiman equation which is consistent with elastic unitarity, analyticity and crossing symmetry.  Using the input from the $\pi\pi$ amplitude, the Khuri-Treiman equation was solved using Pasquier inversion technique. This allowed to establish a significant reduction of the unknown parameters compared to a more straightforward Omn\`es solution. However, the price is the treatment of the left-hand cuts, which is in general not known. We assume, that the unitarity in the physical region, where it can be constrained by the data, plays the key role and does not depend on an accurate form of the unphysical left-hand cuts. The latter we absorbed in the subtraction constants \cite{Guo:2014vya}. With these model assumptions we were able to describe the data from KLOE \cite{Anastasi:2016qvh} and WASA-at-COSY \cite{Adlarson:2014aks} with a minimal number fitting parameters.

The new results are \mbox{$\alpha= -0.025 \pm 0.004$} and
\mbox{$Q=21.6 \pm 1.1$}. Since the experimental data on $\eta\rightarrow 3\pi$ Dalitz plot is very precise now, the main experimental uncertainties come from $I=0$ two-pion scattering amplitudes and the decay width $\Gamma_{\eta\to\pi^+\pi^-\pi^0}$. Improving them are relevant for further $Q$-value and $\alpha$ determinations. 

After submission of our manuscript an improved dispersive analysis based on Omn\`es functions was announced in \cite{Colangelo:2016jmc}. The new $Q$-value is $Q=22.0\pm 0.7$ which is consistent with our estimate.

The codes employed to compute the partial wave amplitudes and the Dalitz plot distribution are available for downloading as well as in an interactive form online at the Joint Physics Analysis Center (JPAC) webpage \cite{website}.

\label{SectionIV}

\begin{acknowledgments}
We thank Astrid Blin for the valuable comments on this manuscript.
This material is based upon work supported in part by the U.S. Department of Energy, Office of Science, Office of Nuclear Physics under contracts  DE-AC05-06OR23177, DE-FG0287ER40365, National Science Foundation under Grants PHY-1415459 and PHY-1205019. The work of I.V.D. is supported by the Deutsche Forschungsgemeinschaft (DFG) through the Collaborative Research Center SFB 1044.  P.G. acknowledges support from Department of Physics and Engineering, California State University, Bakersfield, CA. C.F.-R. work is supported in part by CONACYT (Mexico) under grant No. 251817

\end{acknowledgments}

\bibliographystyle{prsty}
\bibliography{Eta_paper_short}

\begin{thebibliography}{10}

\bibitem{PDG-2015}
K.~A. Olive {\it et~al.}, Chin. Phys. {\bf C38},  090001  (2014).

\bibitem{Aaij:2014jqa}
R. Aaij {\it et~al.}, Phys. Rev. Lett. {\bf 112},  222002  (2014).

\bibitem{Swanson:2006st}
E.~S. Swanson, Phys. Rept. {\bf 429},  243  (2006).

\bibitem{Aaij:2013sfa}
R. Aaij {\it et~al.}, Phys. Rev. Lett. {\bf 111},  101801  (2013).

\bibitem{Aaij:2013bla}
R. Aaij {\it et~al.}, Phys. Rev. Lett. {\bf 112},  011801  (2014).

\bibitem{Battaglieri:2010zza}
M. Battaglieri, Int.J.Mod.Phys. {\bf E19},  837  (2010).

\bibitem{Eugenio:2003}
P. Eugenio, JLAB-E04-005  (2003).

\bibitem{Adolph:2014rpp}
C. Adolph {\it et~al.}, Phys.Lett. {\bf B740},  303  (2015).

\bibitem{Ablikim:2015cmz}
M. Ablikim {\it et~al.}, Phys. Rev. {\bf D92},  012014  (2015).

\bibitem{Sutherland:1966zz}
D.~G. Sutherland, Phys.Lett. {\bf 23},  384  (1966).

\bibitem{Bell:1996mi}
J.~S. Bell and D.~G. Sutherland, Nucl.Phys. {\bf B4},  315  (1968).

\bibitem{Ditsche:2008cq}
C. Ditsche, B. Kubis, and U.-G. Meissner, Eur.Phys.J. {\bf C60},  83  (2009).

\bibitem{Gormley:1970qz}
M. Gormley {\it et~al.}, Phys.Rev. {\bf D2},  501  (1970).

\bibitem{Layter:1973ti}
J.~G. Layter {\it et~al.}, Phys.Rev. {\bf D7},  2565  (1973).

\bibitem{Abele:1998yj}
A. Abele {\it et~al.}, Phys.Lett. {\bf B417},  197  (1998).

\bibitem{Ambrosino:2008ht}
F. Ambrosino {\it et~al.}, JHEP {\bf 0805},  006  (2008).

\bibitem{Adlarson:2014aks}
P. Adlarson {\it et~al.}, Phys.Rev. {\bf C90},  045207  (2014).

\bibitem{Anastasi:2016qvh}
A. Anastasi {\it et~al.}, JHEP {\bf 05},  019  (2016).

\bibitem{Cronin:1967jq}
J.~A. Cronin, Phys.Rev. {\bf 161},  1483  (1967).

\bibitem{Osborn:1970nn}
H. Osborn and D.~J. Wallace, Nucl.Phys. {\bf B20},  23  (1970).

\bibitem{Gasser:1984pr}
J. Gasser and H. Leutwyler, Nucl.Phys. {\bf B250},  539  (1985).

\bibitem{Bijnens:2007pr}
J. Bijnens and K. Ghorbani, JHEP {\bf 0711},  030  (2007).

\bibitem{Colangelo:2009db}
G. Colangelo, S. Lanz, and E. Passemar, PoS {\bf CD09},  047  (2009).

\bibitem{Lanz:2013ku}
S. Lanz, PoS {\bf CD12},  007  (2013).

\bibitem{Schneider:2010hs}
S.~P. Schneider, B. Kubis, and C. Ditsche, JHEP {\bf 1102},  028  (2011).

\bibitem{Kampf:2011wr}
K. Kampf, M. Knecht, J. Novotny, and M. Zdrahal, Phys.Rev. {\bf D84},  114015
  (2011).

\bibitem{Descotes-Genon:2014tla}
S. Descotes-Genon and B. Moussallam, Eur. Phys. J. {\bf C74},  2946  (2014).

\bibitem{Guo:2015zqa}
P. Guo {\it et~al.}, Phys. Rev. {\bf D92},  054016  (2015).

\bibitem{Khuri:1960kt}
N.~N. Khuri and S.~B. Treiman, Phys. Rev. {\bf 119},  1115  (1960).

\bibitem{Kambor:1995yc}
J. Kambor, C. Wiesendanger, and D. Wyler, Nucl.Phys. {\bf B465},  215  (1996).

\bibitem{Anisovich:1996tx}
A.~V. Anisovich and H. Leutwyler, Phys.Lett. {\bf B375},  335  (1996).

\bibitem{Bronzan:1963kt}
J.~B. Bronzan and C. Kacser, Phys. Rev. {\bf 132},  2703  (1963).

\bibitem{Aitchison:1965kt}
I. Aitchison, II Nuovo Cimento {\bf 35},  434  (1965).

\bibitem{Aitchison:1965dt}
I. Aitchison, Physical Review {\bf 137},  B1070  (1965).

\bibitem{Aitchison:1966kt}
I.~J.~R. Aitchison and R. Pasquier, Phys. Rev. {\bf 152},  1274  (1966).

\bibitem{Pasquier:1968kt}
R. Pasquier and J.~Y. Pasquier, Phys. Rev. {\bf 170},  1294  (1968).

\bibitem{Pasquier:1969dt}
R. Pasquier and J.~Y. Pasquier, Phys.Rev. {\bf 177},  2482  (1969).

\bibitem{Stern:1993rg}
J. Stern, H. Sazdjian, and N.~H. Fuchs, Phys.Rev. {\bf D47},  3814  (1993).

\bibitem{Knecht:1995tr}
M. Knecht, B. Moussallam, J. Stern, and N.~H. Fuchs, Nucl.Phys. {\bf B457},
  513  (1995).

\bibitem{Lutz:2011xc}
M.~F.~M. Lutz and I. Vidana, Eur. Phys. J. {\bf A48},  124  (2012).

\bibitem{Gribov:1962fu}
V. Gribov, V. Anisovich, and A. Anselm, Sov.Phys.JETP {\bf 15},  159  (1962).

\bibitem{PhysRev.132.2712}
C. Kacser, Phys. Rev. {\bf 132},  2712  (1963).

\bibitem{Guo:2014vya}
P. Guo, I.~V. Danilkin, and A.~P. Szczepaniak, Eur. Phys. J. {\bf A51},  135
  (2015).

\bibitem{Niecknig:2012sj}
F. Niecknig, B. Kubis, and S.~P. Schneider, Eur.Phys.J. {\bf C72},  2014
  (2012).

\bibitem{Danilkin:2014cra}
I.~V. Danilkin {\it et~al.}, Phys. Rev. {\bf D91},  094029  (2015).

\bibitem{Colangelo:2015kha}
G. Colangelo, E. Passemar, and P. Stoffer, Eur. Phys. J. {\bf C75},  172
  (2015).

\bibitem{Aitchison:1978pw}
I.~J.~R. Aitchison and J.~J. Brehm, Phys.Rev. {\bf D17},  3072  (1978).

\bibitem{Guo:2014mpp}
P. Guo, Phys. Rev. {\bf D91},  076012  (2015).

\bibitem{Guo:2015kla}
P. Guo, Mod. Phys. Lett. {\bf A31},  1650058  (2015).

\bibitem{Amoros:2001cp}
G. Amoros, J. Bijnens, and P. Talavera, Nucl.Phys. {\bf B602},  87  (2001).

\bibitem{Aoki:2016frl}
S. Aoki {\it et~al.}, Eur. Phys. J. {\bf C77},  112  (2017).

\bibitem{GarciaMartin:2011cn}
R. Garcia-Martin {\it et~al.}, Phys.Rev. {\bf D83},  074004  (2011).

\bibitem{Colangelo:2016jmc}
G. Colangelo, S. Lanz, H. Leutwyler, and E. Passemar, Phys. Rev. Lett. {\bf
  118},  022001  (2017).

\bibitem{website}
\url{http://www.indiana.edu/~jpac/index.html}  .

\end{thebibliography}

\end{document}